\documentclass{aa}

\usepackage{ulem}
\usepackage{txfonts}
\usepackage{graphicx}

\usepackage{natbib}
\bibpunct{(}{)}{;}{a}{}{,} 

\begin{document}

\title{De-biasing interferometric visibilities in VLTI-AMBER data of low SNR observations
        \thanks{Based on VLTI-AMBER observations} }

\author{    G. Li Causi \inst{1},
            S. Antoniucci \inst{1,2}
            \and
            E. Tatulli \inst{3,4}
            }

\institute{INAF-Osservatorio Astronomico di Roma, Via Frascati 33, I-00040 Monteporzio Catone
           \and
           Universit\`a degli Studi di Roma `Tor Vergata', via della Ricerca Scientifica 1, I-00133 Roma
           \and
           INAF-Osservatorio Astrofisico di Arcetri, Largo E. Fermi 5, I-50125 Firenze
           \and
           Laboratoire d'Astrophysique Observatoire de Grenoble, BP 53 F-38041, GRENOBLE Cédex 9
           }

\offprints{Gianluca Li Causi, email address: licausi@oa-roma.inaf.it}

\date{Received date / Accepted date}

\abstract {}
{   
    We have found that the interferometric visibilitiesof VLTI-AMBER observations,
    extracted via the standard reduction
    package, are significantly biased when faint targets are concerned.
    The visibility biases derive from a time variable fringing
    effect (correlated noise) appearing on the detector.
}
{   
    We have developed a method to correct this bias that consists
    in a subtraction of the extra power due to such correlated noise,
    so that the real power spectrum at the spatial frequencies of the
    fringing artifact can be restored.
}
{   
    This pre-processing procedure is implemented in
    a software, called \texttt{AMDC} and available to the community, to be run
    before the standard reduction package.
    Results obtained on simulated and real observations
    are presented and discussed.
} {}


\titlerunning{AMBER visibilities from low SNR observations}
\authorrunning{Li~Causi, Antoniucci, Tatulli}
\maketitle

\section{Introduction}\label{Introduction}
Astronomical interferometry consists in combining in phase the light from two or more telescopes, to
produce an interferogram at the focal plane, i.e. an image composed of several fringes, which is
recorded by the detector. For such interferometric data the presence of a common readout electronic
effect known as \textit{fringing}, normally negligible in single telescope observations, is very
dangerous.

The term \textit{fringing} is commonly used to indicate a spurious pattern composed of parallel
fringes overlaid on the image. It can be stable or variable with time, both in phase and amplitude,
sometimes also variable in period and orientation, or it can be even non-periodic, so that it is
configurable as a structured random noise, also called as \textit{correlated noise}\footnote{In
astronomical imaging there is another effect known as \textit{fringing}, where irregular fringes
appear on the sky background. This is caused by the light of the sky emission lines that interferes
within the faces of the detector window. We will not refer to this particular fringing effect in this
work.}.

Considering such fringing instead of the true interferometric signal may lead to wrong visibility
estimates, especially when dealing with low SNR (Signal to Noise Ratio) observations.

We have discovered the presence of this correlated noise effect in many observations of the
interferometric instrument VLTI/AMBER (\citealt{Petrov_AMBER}), especially those characterized by a
very low SNR: in these data, the main fringing component happens to have the same period as the true
interferogram, thus biasing the visibility extracted by the standard AMBER processing package
\textit{amdlib}.

Actually, the design of AMBER and its data reduction software have been conceived taking into account
the presence of the FINITO fringe tracker \citep{FINITO}, i.e. with the assumption of stable fringes
ensuring long integration times and photon-noise regime. Because of problems related to structural
vibrations within the VLTI, FINITO is not yet available for observations with the Unitary Telescopes
(UTs) and this explains why, even for faint sources, AMBER is currently used with short integration
times that often force the instrument to work at fringe SNR close to one. In such a situation, the
science signal is basically comparable to the fringing, so that the correlated noise cannot be
neglected anymore in the data processing.

We present here a method that we have developed to remove the bias introduced by the correlated noise
to the VLTI-AMBER visiblities.

After a short presentation of the AMBER instrument (sec.~\ref{The AMBER instrument}), we present a
brief report of the standard data reduction procedures (sec.~\ref{Observations and data reduction
with AMBER}); then, we describe the fringing artifact (sec.~\ref{The detector artifact}) and the
method we have developed to remove it (sec.~\ref{removing_the_artifact}). Finally, we discuss the
results obtained applying the correction on real and simulated data
(sec.~\ref{corrected_visibilities}).

\section{The AMBER instrument}\label{The AMBER instrument}

\begin{figure}[!t]
    \begin{center}
    \includegraphics[width=7cm]{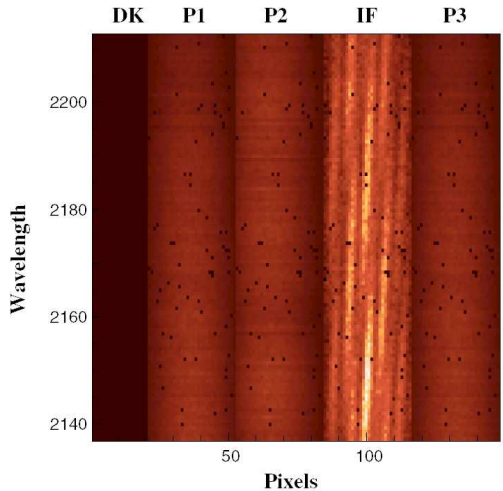}
    \caption{The five channels of an AMBER image: from left to right the darkened channel DK, which
    is a physically masked region of the detector; the photometries P1 and P2; the interferogram IF;
    the photometry P3 (adapted from \citealt{Tatulli_AMBER}).}
    \label{Fig_AMBER_channels}
    \end{center}
    \end{figure}

AMBER, the first-generation near-infrared interferometric instrument in operation at the ESO Very
Large Telescope Interferometer (VLTI), has a beam combiner fed by three fibers, capable to
simultaneously provide multi-wavelength visibilities and closure phases (\citealt{Petrov_AMBER}).
AMBER works combining the three beams on the same area of an IR array, thus forming three
superimposed fringe patterns of known spatial periods (\citealt{Tatulli_AMBER}). This approach,
called \textit{multiaxial combination} allows the information of both the three beams photometry and
the three correlated fluxes to be encoded in a small detector area. Spectral dispersion is added
along the orthogonal direction, with a spectral window which depends on the resolution, selected
among R=30~(LR-mode), R=1500~(MR-mode) and R=12000~(HR-mode). This results in frames which are
divided in five vertical regions or ``channels'', as shown in Fig.~\ref{Fig_AMBER_channels}, where
the OPD (Optical Path Difference) is along the x-axis and wavelengths along the y-axis. Three
channels carry the photometric information, one for each telescope, while the interferometric channel
contains the overlay of the three interferograms, related to the three baselines; these are
characterized by different fringe periods, orientations and shapes, which are extracted by a custom
data reduction algorithm (see sec.~\ref{Observations and data reduction with AMBER}).

Due to the complex optical train, to the spectral dispersion and mainly to the OPD instabilities
caused by the atmospheric piston variations (i.e. the zero-order phase variations over different
telescopes), AMBER is limited to a K-band magnitude from 4.5 to 5.5, in MR-mode with the UTs,
depending on seeing conditions and adaptive optics performance, and the single frame integration time
(DIT) is restricted to a range from 25 to 100~ms; the fringe-tracker FINITO (\citealt{FINITO}) is
foreseen in the next future to overcome these limits\footnote{Note that FINITO is already available
with the Auxiliary Telescopes}.

\section{Observations and data reduction with AMBER}\label{Observations and data reduction with AMBER}

We have analyzed several AMBER observations (especially the ones taken in medium spectral resolution
mode, where the number of photons per spectral channel is drastically reduced) and all of them
display some correlated noise fringing. In order to illustrate the correlated noise problem and the
correction procedure, we will refer here to the observations of the young stellar object Z~CMa~A
(\citealt{koresko91}), which are characterized by a very low fringe SNR and by a strong noise
artifact. The target (a source with a K~band magnitude slightly variable around 4.2~mag) has been
observed using the VLTI UT1-UT2-UT4 baselines in MR-mode (R=1500), with a spectral window centered at
$2.16~\mu$m and using a DIT of $50$~ms.

\begin{figure*}[!t]
    \begin{center}
    \includegraphics[width=14cm]{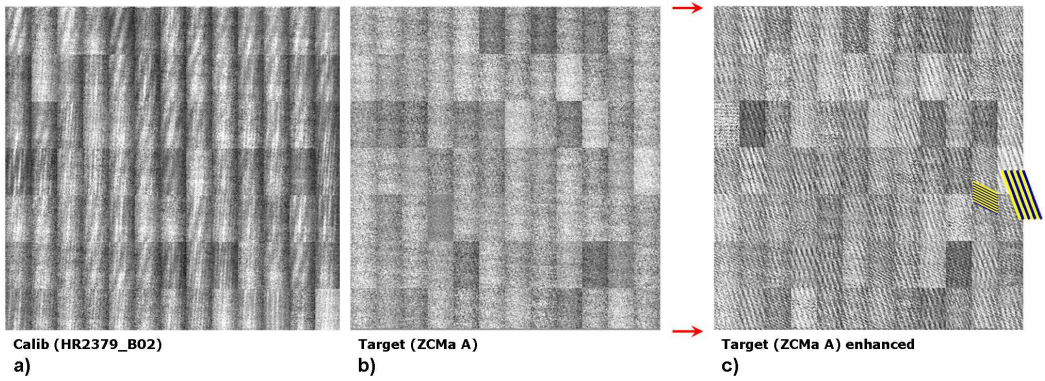}
    \caption{Time sequence of the interferometric channel for the calibrator HR2379 (a)
    and for the target Z~CMa~A (b). Panel c): enhancement of the same sequence
    for the target (see footnote~\ref{footnote_enhancement}):
    at least two fringe systems are visible, as indicated.}
    \label{Fig_frames_calib_vs_ZCMa}
    \end{center}
    \end{figure*}

A part of the frame sequence from the observations is reported in
Fig.~\ref{Fig_frames_calib_vs_ZCMa}: panel \textit{a} displays measurements on the calibrator
(HR2379, diameter = $1.75$mas, K-band magnitude = $2.24$), while panel \textit{b} shows frames
acquired on the target Z~CMa~A . We can note by direct eye inspection that interferometric fringes
appear on the calibrator frames, while no fringes seem to be present on the target observations.


In order to derive the interferometric visibilities we used the standard AMBER reduction pipeline
\textit{amdlib}, described in detail by \citet{Tatulli_AMBER}. The procedure used in the pipeline is
based on the so called ``P2VM matrix'' method (\citealt{Millour_P2VM}), in which the three baseline
fringe patterns are disentangled by solving a linear system describing the fringe packets for each
spectral channel separately.

Following the standard data reduction procedure, we get calibrated visibilities of $1.26\,\pm\,0.08$,
$0.88\,\pm\,0.04$ and $0.83\,\pm\,0.04$ for the three baselines, respectively. These non-zero values
appear in contrast with the apparent lack of fringes in the frame sequence shown in
Fig.~\ref{Fig_frames_calib_vs_ZCMa}~panel~\textit{b}, but using a proper visual enhancement (see
footnote~\ref{footnote_enhancement}) at least two fringe systems can be detected in the target
observations (Fig.~\ref{Fig_frames_calib_vs_ZCMa}, panel \textit{c}). However, the visibility
measured on the first baseline is greater than one, while by definition the normalized visibility
should be comprised between 0 and 1. Two main causes can be invoked to explain this inconsistent
result: \textit{(i)} a change of atmospheric conditions between the observation of the science target
and the calibrator, which prevents us from properly correcting the visibility for the atmospheric
transfer function, or \textit{(ii)} the presence of detector artifacts which bias the visibility
estimate. In the next section, we show that this latter effect is indeed present in the Z~CMa~A data
and that this situation is commonly found in all AMBER low SNR observations.

\section{The detector artifact}\label{The detector artifact}

\begin{figure*}[!t]
    \begin{center}
    \includegraphics[width=14cm]{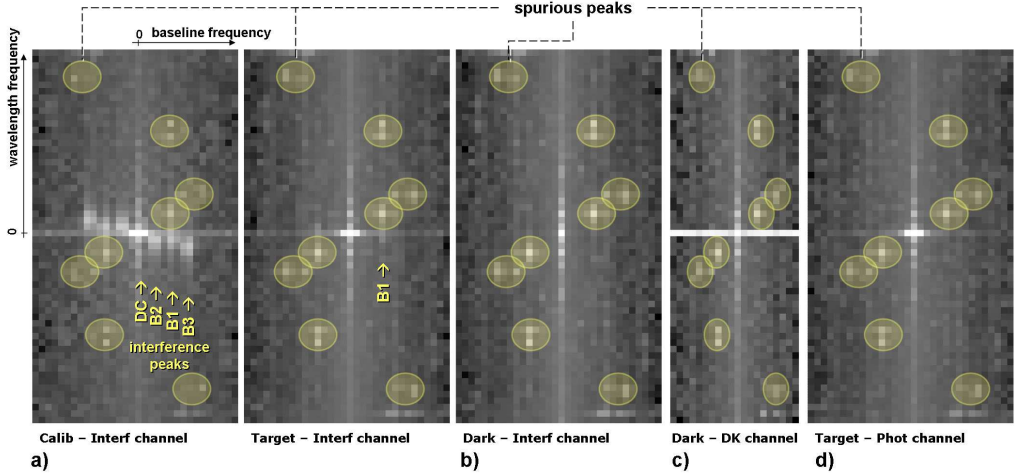}
    \caption{Panel a): comparison between the average power spectra of the interference channel from the
    observations of the calibrator and of the target. The origin of the frequency plane
    is at the center, with baselines frequencies along the x-axis and wavelength frequencies
    along the y-axis. The three baselines peaks are visible in the calibrator
    (\texttt{B2} - \texttt{B1} - \texttt{B3} labels), while only baseline~1 signal is
    clearly visible in the target. Some spurious power peaks can be identified spread over
    the plane (circles), which appear at the same positions for both the calibrator and target.
    Noticeably, they also appear in the dark observations (panel b), in the masked channel DK
    (panel c) and in the photometric channels (panel d), where no fringes at all are expected.}
    \label{Fig_power_spectra_Calib_ZCMa_dark}
    \end{center}
    \end{figure*}

In order to investigate the cause of the problematic results described above, we performed a 2-D
Fourier analysis of the frames. Looking at the averaged power spectrum (PS) of the interference
channels for the calibrator and the target (Fig.~\ref{Fig_power_spectra_Calib_ZCMa_dark}, panel a) we
easily recognize the three baselines peaks plus the continuum peak at zero frequency, but we also
notice some extra power peaks in other locations, as indicated in the figure. Remarkably, we found
exactly the same extra peaks both in the photometric channels and in all the channels of the dark
observations, where no fringes at all are expected (last three panels in
Fig.~\ref{Fig_power_spectra_Calib_ZCMa_dark}). Even the PS of the physically masked channel DK shows
the same peaks, which means that they are not due to light interference but are spurious artifacts
coming from the detector electronics or read-out process. The most relevant issue about these
artifacts is that the brightest peaks happen to lie at the same spatial frequencies of baseline~1 and
partly of baseline~3. So, even in absence of light on the detector, there are some artificial
``detector fringe patterns'' in the AMBER frames, which have the same spatial periods of the
baselines fringes.

The \textit{amdlib} software cannot distinguish this fringing from the true baseline interferograms,
because its analysis is based on line-by-line fringe fitting, allowing for random fringe orientation
as induced by the atmospheric piston variation. This is the reason why we get wrong visibilities from
the \textit{amdlib} extraction: the algorithm fits the fringing artifact instead of the true signal
because this latter, in case of low SNR targets, has equal or smaller amplitude than the former.

We point out that this pattern is not removed by the normal dark subtraction operated by
\textit{amdlib} because it is rapidly variable in time and thus has a different phase shift in each
frame. Consequently, this spurious pattern is not present in the average dark frame and the dark
subtraction is not able to remove it.

\begin{figure}[!t]
    \begin{center}
    \includegraphics[width=9cm]{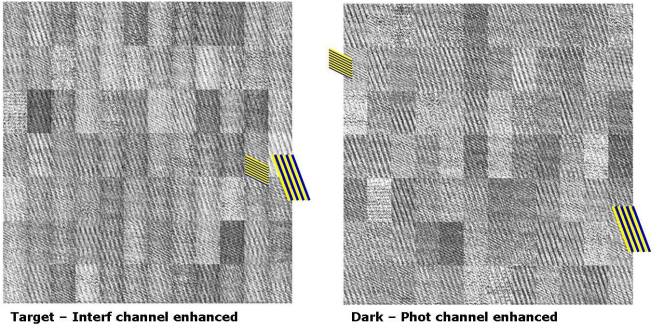}
    \caption{Comparison between the fringe systems observed in the interference channel
    of the target and the ones observed in a photometric channel of the dark observations
    (enhanced, see footnote~\ref{footnote_enhancement}): both of them are artifacts of the AMBER detector.}
    \label{Fig_fringes_on_dark_and_ZCMa}
    \end{center}
    \end{figure}

Fig.~\ref{Fig_fringes_on_dark_and_ZCMa} shows a direct comparison between the frames of the target
and the frames of a photometric channel of the dark (where we would expect no fringes at all): we
clearly see exactly the same fringe systems, thus confirming that our Z~CMa~A observations are
dominated by the correlated noise and that all the fringes displayed in
Fig.~\ref{Fig_frames_calib_vs_ZCMa}, panel~c, are spurious.

\section{Fringing removal: the \texttt{AMDC} software}\label{removing_the_artifact}

The fringing affecting AMBER data is likely caused by some electromagnetic interferences (EMI)
corrupting the detector read-out process (see e.g. \citealt{Mardones} and \citealt{Beckmann}).
Hereafter, we present our approach to correct this problem via sofware.

The procedure we have developed is implemented in the \texttt{AMDC} tool, which stands for ``AMBER
Detector Cleaner''. AMDC effectively removes the fringing artifact from AMBER raw frames, thus
allowing users to properly reduce observations obtained in poor instrumental conditions.

The program is written in IDL (Interactive Data Language) and is downloadable at
\texttt{http://www.mporzio.astro.it/$\sim$licausi/AMDC/}.

It works as a pre-processing tool to be run on the original AMBER frames before the standard
\textit{amdlib} reduction is performed. It creates a copy of the original data set where the fringing
is no longer present, so that the standard processing can be performed on these corrected data.

Our approach consists in \textit{(i)} marking the locations of the artifacts in the average PS, then
\textit{(ii)} replacing a modelled PS estimate within the marked frequency regions on a
frame-by-frame basis and finally \textit{(iii)} rebuilding the image frames via inverse Fourier
Transform.

Working on power spectra, this method will effectively subtract out the contribution to the power due
to the correlated noise in all the involved frequencies, but cannot restore their original phases,
for the reasons we will explain below. Nonetheless, the reconstruction error of both the modulus and
phase of the final visibility is negligible providing that the artifacts peaks do not overlap with
the baseline peaks in the PS. This situation that can be easily avoided by setting a suitable
non-zero OPD during the fringe centering stage at the beginning of the AMBER observational procedures
(in fact, a variation of the central OPD causes the science peaks to shift vertically in the
frequency plane, thus placing them away from the artifact regions). On the contrary, in the unlucky
case where artifact and interferometric peaks do overlap in the PS, the algorithm will still allow
\textit{amdlib} to restore the correct average $V^2$, but not the single visibilities $V_i$ of each
frame.

From a mathematical point of view, we proceed as follows. At first we build a binary mask to mark the
frequencies affected by the artifacts, which are identified on the average PS of the dark frames by
an algorithm which repeats three times a sigma-clipping followed by a morphological dilation
operator, so as to mark both the main peaks and their skirts.

For each photometric channel of the dark observations ($D_{p,i}$), we compute the average (over the
$n$ frames) power $\widehat{D}_{p}$:

    \begin{equation}
    \lefteqn{\widehat{D}_p~=~\frac{\sum_i^n|FT(D_{p,i})|^2}{n}}
    \end{equation}

Then we derive the binary mask $M_p$ marking the frequencies where the power differs from the median
by more than a threshold of $K$ standard deviations. The median is computed on a column-by-column
basis, given the vertical symmetry of the PS.

    \begin{equation}\label{eq_mask}
    \lefteqn{M_p(\nu_x, \{\nu_y: \widehat{D'}_p(\nu_x, \nu_y) > [K * \sigma_{\widehat{D'}_p(\nu_x, *)}] \} ) ~=~ 1,~0~elsewhere }
    \end{equation}

\noindent where

    \begin{equation}
    \lefteqn{\widehat{D'}_p(\nu_x, *)~=~\widehat{D}_p(\nu_x, *)~-~MEDIAN(\widehat{D}_p(\nu_x, *))}
    \end{equation}

\noindent and $\sigma$ is a robust standard deviation estimator. Using this procedure we mark the
power peaks, but we also need to mark their wings. Thus, we apply again equation~\ref{eq_mask} using
a lower threshold $K\sigma$, but only in the pixels surrounding the spurious peaks we have previously
identified, then iterating this procedure if needed.

%

\noindent Finally, the masks $M_p$ computed for each photometric channel are median-averaged to build
a single mask $\overline{M}$ that we will use for all the channels.

\begin{figure}[!t]
    \begin{center}
    \includegraphics[width=8cm]{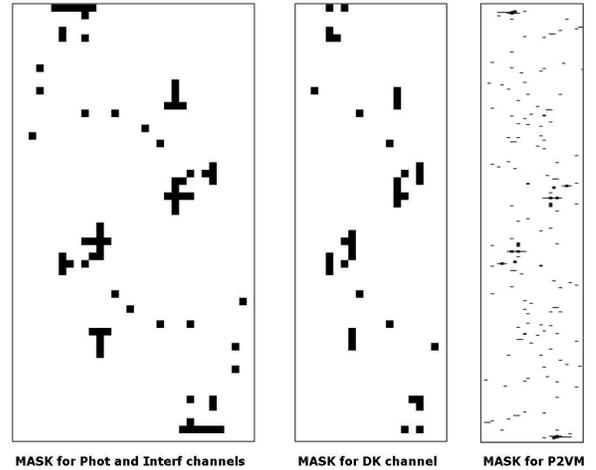}
    \caption{Examples of masks, computed from the average power spectrum of the dark frames, used
    to remove the detector pattern artifact from the interferometric and photometric channels (left),
    from the masked channel DK (middle) and from the P2VM frames (right).}
    \label{Fig_power_masks}
    \end{center}
    \end{figure}

The above procedure is repeated for each type of AMBER observation files (calibrator, target, P2VM
calibration, spectral calibration) in order to produce the right mask to be used for each of them
(Fig.~\ref{Fig_power_masks}).

Analyzing the PS of the single frames we also notice that the detector pattern is characterized by a
highly variable amplitude, so that some frames appear heavily affected by artifact, while others do
not show any bias. This means that we cannot simply subtract the average PS of the dark frames from
the PS of each target frame; instead we need an estimate of the instantaneous artifact PS. To get
this estimate we assume that the spurious power in the interferometric channel of a frame is given by
the average of the spurious power in the photometric channels of the same frame. We obtain this
latter as the difference between the currently measured PS and a model of the uncorrupted PS, $W_p$.
This model is built from the average PS of the P2VM files where only one photometric channel at a
time is lit \citep{Tatulli_AMBER}. Indeed, in these files the other channels are darkened, so that
their PS only contain the simultaneous spurious peaks and we subtract it from the PS of the opened
channel. This way we get the three perfectly cleaned photometric PS to be used as $W_p$ models.


In formulae, since these models and the observed PS of the photometric channels of the target
$\widehat{T}_{p,i}$ differ for a global constant and a scaling factor, we adapt the model to the
measured PS by a linear fit, restricted to the uncorrupted frequencies, i.e. where
$\overline{M}(\nu_x,\nu_y)=0$:

    \begin{equation}
    \lefteqn{|[a_0 * W_p + a_1] - \widehat{T}_{p,i}|^2 < \varepsilon}
    \end{equation}

\noindent where $a_j$ are the coefficients of the fit.

\noindent Thus, the estimates of the instantaneous artifact power $A_{p,i}$ on the masked frequencies
(where $\overline{M}=1$) are given by:

    \begin{equation}
    \lefteqn{A_{p,i}~=~\widehat{T}_{p,i} - \overline{W}_p}
    \end{equation}

\noindent and zero elsewhere, where $\overline{W}_p$ are the best fit models.

\noindent Finally, we derive a cleaned PS of each frame by directly subtracting the $A_{p,i}$ from
the photometric channels $p$ and their median-average $\overline{A}_{p,I}$ from the interferometric
channel $I$:
    \begin{eqnarray}
    \lefteqn{\widehat{T}^{Corr}_{p,i}=\widehat{T}_{p,i} - A_{p,i}} \\
    \lefteqn{\widehat{T}^{Corr}_{I,i}=\widehat{T}_{I,i} - \overline{A}_{p,i}}
    \end{eqnarray}

\noindent In summary, we are subtracting from each frame the power of the detector noise present at
the moment of the frame readout.

\begin{figure*}[!th]
    \begin{center}
    \includegraphics[width=12cm]{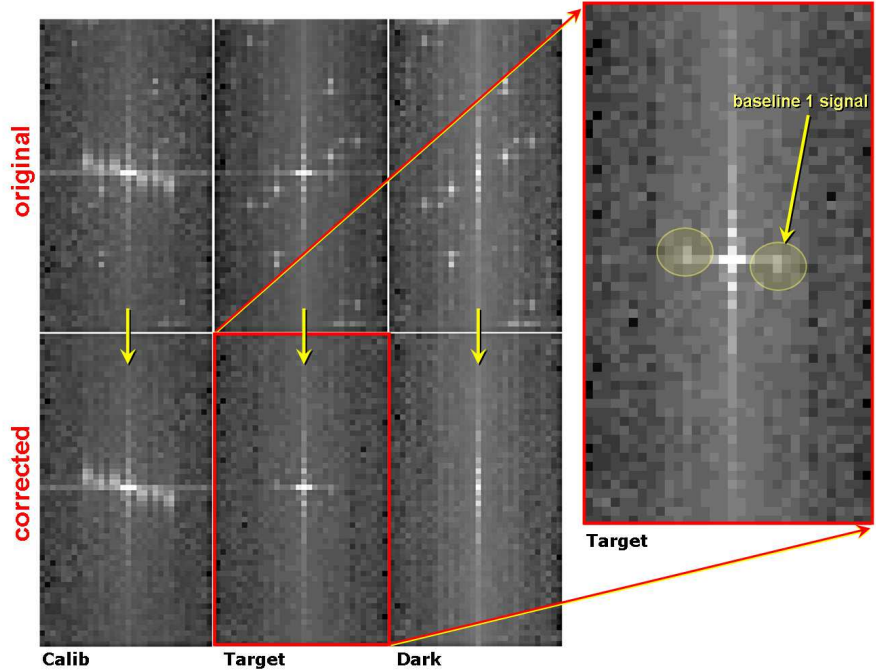}
    \caption{Removal of the artifact peaks from the power spectra of calibrator (left),
    target (middle) and dark frames (right): original power images are displayed on the top
    row and corrected ones at bottom. The faint interferometric signal in baseline~1 of our target
    is now clearly visible (zoomed panel).}
    \label{Fig_power_masked}
    \end{center}
    \end{figure*}

Fig.~\ref{Fig_power_masked} shows the result of this process applied to our data: we note that
\textit{(i)} the only remaining peaks in the PS of the calibrator are the three baseline signals,
\textit{(ii)} no spurious peaks remains in the PS of the corrected dark frames and \textit{(iii)} in
the target PS we clearly see the clean peaks of the first baseline, which carries the dominant
interferometric signal of these data (zoomed panel in the figure).

A thorough analysis has shown that it is not possible to recover also the uncorrupted phase
information $\varphi_{I,i}(\nu_x,\nu_y)$, unless each pixel is tagged with a precise read-time stamp,
which is not available in the AMBER files. Indeed, the external EMI $E(t)$ combines with the complex
pixels readout temporal sequence $U(t)$ producing a different spatial modulation $V(x,y)$ for each
frame readout. In the AMBER case, the detector is read by a double correlated sampling, which follows
a ``Reset---Non~Destructive~Read---Destructive~Read'' scheme, where the two readout sequences have
different timings and only the difference between them within the channel windows are transferred.
This implies that the final fringing pattern has a spectrum of spatial frequencies with a phase
distribution arbitrarily complex, even if the EMI is monochromatic. This fact prevents us from
recovering the phased frequency spectra at the artifact frequencies, so affecting the single frame
visibilities in case of coincidence with baselines frequencies. This is the reason why a suitable OPD
correction must be adopted, as already said, to avoid any possible overlapping of baseline and
artifact signals in the PS.

\noindent At the end of the process the reconstructed image frames, that no longer contain the
detector artifact, are computed by means of an inverse Fourier Transform and put in place of the
original (corrupted) frames in the AMBER file:
    \begin{eqnarray}
    \lefteqn{T^{Corr}_{I,i}~=~FT^{-1}([\widehat{T}^{Corr}_{I,i}, \varphi_{I,i}])}  \\
    \lefteqn{T^{Corr}_{p,i}~=~FT^{-1}([\widehat{T}^{Corr}_{p,i}, \varphi_{p,i}]).}
    \end{eqnarray}

\noindent where the phases $\varphi_{*,i}(\nu_x,\nu_y)$ are the original ones.

\begin{figure}[!th]
    \begin{center}
    \includegraphics[width=9cm]{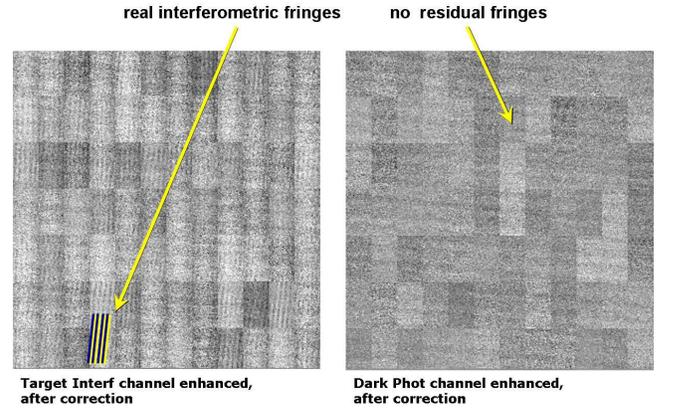}
    \caption{\texttt{AMDC} enhancement of the frame sequence after removal of the detector artifact:
    the true interferometric fringes now appear on the target (left), while no residual fringes
    remain on the dark frames (right). Compare them to Fig.~\ref{Fig_fringes_on_dark_and_ZCMa}.}
    \label{Fig_frames_corretti}
    \end{center}
    \end{figure}

\noindent Fig.~\ref{Fig_frames_corretti} shows the result of the \texttt{AMDC} correction on the
frame sequence of Z~CMa~A: we clearly see\footnote{Note: for this figure, as for some previous ones
displaying frame sequences, we have used the \texttt{AMDC} enhancement function because of the very
low contrast of the fringes, hardly visible even on the screen; this is obtained by magnifying, by
one order of magnitude, the power at the frequencies identified by the mask, or that of the baselines
peaks.


\noindent Note as well  that an ``eye-inspection'' of the frames is in general suitable just for a
first quick and qualitative analysis of the data, since fringes might be present without being strong
enough to be seen above the noise level on the screen, as it happens here for the second and third
baselines of Z~CMa~A. As a matter of fact, a quantitative parameter, namely the SNR fringe contrast
as defined in Eq.~(20) of \citealt{Tatulli_AMBER}, is the criterium to be used for a thorough data
analysis.\label{footnote_enhancement}} that the spurious fringes are removed and the real
interferometric fringes appear, while no residual fringes are visible in the sequence of dark frames,
thus confirming that what we get is not an artifact of our processing.

The correction procedure just described for the target frames must then be applied to all the frames
of the data set, including the spectral calibration files, the P2VM calibration files, the calibrator
files and the dark observations themselves. Such operations have to be done at a pre-processing
stage, before launching the standard \textit{amdlib} sequence \citep{AMBER_DRS}, consisting in
spectral offsets computation, building of P2VM matrix and visibility extraction.

This procedure is performed in a semi-automatic way by \texttt{AMDC}, which also provides some
functions for a first analysis of the data, aimed to find out whether any correlated noise is present
and to better visualize it. Procedures are available for displaying the frames sequence and their
average PS, for automatically or manually defining the binary mask, for applying the artifact
correction creating the restored FITS files, and for enhancing the visualization of the fringing.

%
%
%
%
%
%
%
%
%

\section{Corrected visibilities}\label{corrected_visibilities}

The application of the \texttt{AMDC} correction to Z~CMa~A data provides the results reported in
Table~\ref{Table_Visibilities}, which shows the raw visibilities of target and calibrator before and
after the artifact removal.

\begin{figure}[!t]
    \begin{center}
    \includegraphics[width=10cm]{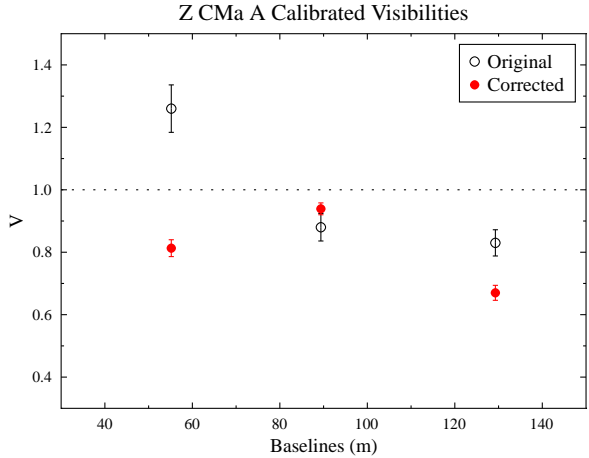}
    \caption{Visibilities of Z~CMa~A before and after the \texttt{AMDC} correction; noticeably,
    no baselines have visibilities greater than one after the correction.}
    \label{Fig_calibrated_visibilities}
    \end{center}
    \end{figure}

The target and the calibrator are not equally affected, as expected for sources with different SNR:
in particular,  we find a relative bias up to 50\% for the uncorrected raw visibilities of Z~Cma~A,
while the results on the calibrator appear to be only slightly corrupted. This explains the
overestimate of the calibrated visibilities of Z~CMa~A that we reported in sec.\ref{Observations and
data reduction with AMBER};  the values we obtain after the fringing removal are 0.81\,$\pm$\,0.03,
0.94\,$\pm$\,0.02 and 0.67\,$\pm$\,0.02, without meaningless values greater than one on the first
baseline (Fig.~\ref{Fig_calibrated_visibilities}).

\begin{table*}[!t]
\caption{Comparison of corrected and uncorrected visibilities. \label{Table_Visibilities}} \small{
\begin{center}
\begin{tabular}{c c l l l l l l}
\hline
  \multicolumn{2}{c}{Baselines}   & \multicolumn{4}{c}{Raw Visibilities}           & \multicolumn{2}{c}{Calibrated Visibilities}\\
\hline
     & u, v    & \multicolumn{2}{c}{Z~CMa~A}    & \multicolumn{2}{c}{HD2379}  & \multicolumn{2}{c}{Z~CMa~A}\\
\#   & (meters)    & Uncorrected  & Corrected       & Uncorrected  & Corrected      & Uncorrected  & Corrected\\
\hline
1    & 24.8,~49.3     & 0.68 $\pm$\,0.04  & 0.45 $\pm$\,0.01     & 0.51 $\pm$\,0.01  & 0.46 $\pm$\,0.01  & 1.26 $\pm$\,0.08  & 0.81 $\pm$\,0.03\\
2    & 88.4,~13.1     & 0.60 $\pm$\,0.01  & 0.62 $\pm$\,0.01     & 0.60 $\pm$\,0.03  & 0.60 $\pm$\,0.03  & 0.88 $\pm$\,0.04  & 0.94 $\pm$\,0.02\\
3    & 113.2,~62.4    & 0.58 $\pm$\,0.02  & 0.58 $\pm$\,0.02     & 0.52 $\pm$\,0.02   & 0.52 $\pm$\,0.02  & 0.83 $\pm$\,0.04  & 0.67 $\pm$\,0.02\\
\hline
\end{tabular}
\end{center}
Note: Raw and calibrated visibilities, averaged over wavelengths, before and after artifact
correction by \texttt{AMDC}, resulting from \textit{amdlib} extraction with 50\% selection by flux
criterium, then 50\% selection by SNR, then all frames binning.}
\end{table*}

Moreover, our analysis has shown that dispersion of the visibility among the observation files is
noticeably reduced after the correction: this also partly explains the large visibility dispersion
normally found in AMBER observations of weak sources, which is therefore not only due to the VLTI
instrumental/atmospheric OPD variations. Unfortunately, we have not been able to evaluate the
\texttt{AMDC} results on the closure phase, given the extremely low SNR of the data and that fringes
must be present in the three baselines at the same time.

\begin{figure}[!t]
    \begin{center}
    \includegraphics[width=10cm]{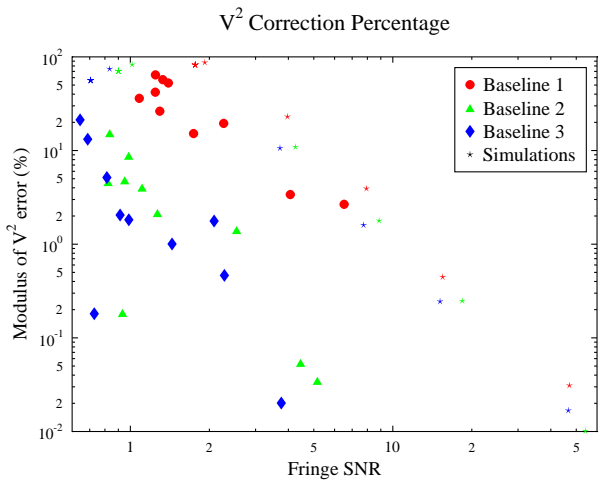}
    \caption{Percentage correction of the instrumental squared visibility on different
    data sets and synthetic simulations: the artifact effect become dominant as detection limit is approached.}
    \label{Fig_percentage_correction}
    \end{center}
    \end{figure}

The same correction procedure has been applied to other AMBER observations of low flux targets
(courteously provided by several users of \texttt{AMDC}) and to simulated data. In all the cases we
have found similar improvements after fringing removal. In Fig.~\ref{Fig_percentage_correction} we
show a plot of the correction percentage of the instrumental visibilities, as a function of the
fringe SNR. We note that the artifact effect tends to increase with decreasing fringe SNR.

We have also found that the locations of the artifact peaks in the frequency plane are very similar
in all the observations we have analyzed; this fact gives an indication of the stability of the
unknown sources of the correlated noise during the time period covered by the observations (from
February~2005 to June~2006).


\section{Conclusions}\label{Conclusion}

While analyzing VLTI-AMBER observations, a fringing artifact (correlated noise on the instrument
detector) has been found on the data frames; this biases the visibility computation for low SNR
targets. This spurious pattern produces a series of spatial frequency peaks in the power spectrum of
the frames, some of which fall at the same baseline frequencies of the interferometric signal, thus
altering the computed visibilities. We have shown that this effect becomes more important in faint
sources, causing the instrumental visibility to be overestimated when dealing with limiting magnitude
targets. We have presented a procedure to restore the un-biased data and we provide to the community
the algorithm of this method implemented in a software called \texttt{AMDC} (freely available at
\texttt{http://www.mporzio.astro.it/$\sim$licausi/AMDC/}), so that AMBER users can perform optimal
data reduction for observations obtained in poor instrumental conditions. AMDC works as a
pre-processing tool to be run on the original AMBER frames before the standard \textit{amdlib}
extraction is performed. The software is planned to be implemented in a future version of
\textit{amdlib}.


\section{Acknowledgements}\label{Acknowledgements}
The authors are very grateful to M.~Benisty, F.~Cusano, A.~Domiciano~de~Souza, G.~Duvert,
O.~Hernandez, A.~Isella, F.~Malbet and W.~de~Wit for providing their observation data sets, which
have been joined to our data for filling in the diagram in Fig.~\ref{Fig_percentage_correction}. We
also thank G.~Duvert for providing very useful numerical simulations that we have used to test our
algorithm and U.~Beckmann for explanations of the AMBER readout.

\bibliographystyle{aa} 
\bibliography{references2} 

\clearpage

\end{document}